# FOSY: Segmental Backbone Assignment in Intrinsically Disordered Proteins

Dmitry M. Lesovoy[b]*, Tatiana Agback[a,d] , Panagiota S. Georgoulia[a], Irena Matečko-Burmann[e], Björn M. Burmann[a,c,f], and Vladislav Y. Orekhov*[a,c]

[a] Prof. V. Y. Orekhov, Dr. T. Agback, Dr. P. S. Georgoulia, Prof. B. M. Burmann
Department of Chemistry and Molecular Biology
University of Gothenburg, Box 465, 405 30, Gothenburg, Sweden
E-mail: vladislav.orekhov@nmr.gu.se

[b] Dr. D. M. Lesovoy
Department of Structural Biology, Shemyakin-Ovchinnikov,
Institute of Bioorganic Chemistry RAS, 117997, Moscow, Russia

[c] Prof. V. Y. Orekhov, Prof. B. M. Burmann
Swedish NMR Centre and SciLifeLab,
University of Gothenburg, Box 465, 405 30, Gothenburg, Sweden

[d] Dr. T. Agback
Department of Molecular Sciences,
Swedish University of Agricultural Sciences, PO Box 7015, 750 07 Uppsala, Sweden.

[e] Dr. I. Matečko-Burmann
Department of Psychiatry and Neurochemistry, Wallenberg Centre for Molecular and Translational Medicine,
University of Gothenburg, Box 465, 405 30, Gothenburg, Sweden

[f] Prof. B. M. Burmann
Wallenberg Centre for Molecular and Translational Medicine,
University of Gothenburg, Box 465, 405 30, Gothenburg, Sweden

**Abstract:**

Backbone resonance assignment is a prerequisite for most biomolecular NMR applications, yet conventional multidimensional strategies frequently fail for intrinsically disordered proteins (IDPs) and regions (IDRs) because of severe spectral overlap, rapid amide proton exchange with water, and missing sequential correlations. In many biological applications, however, complete protein assignment is unnecessary, as only a limited sequence segment surrounding a functional site is required. Here we introduce segmental backbone assignment, an assignment strategy implemented by FOcused SpectroscopY (FOSY), which concentrates experimental effort on relatively short regions while retaining the high-dimensional sequential connectivity needed for unambiguous assignments. We present a self-consistent suite of selective two-dimensional FOSY experiments that enables bidirectional assignment walks along the protein sequence through complementary forward and backward transfer schemes. The methodology employs frequency-selective polarization transfer to replace high-dimensional experiments with sensitive and readily interpretable 2D spectra while preserving the information content of multidimensional correlation experiments. The approach is demonstrated by completing the assignment of G302–K311 segment, which is missing in the published assignment of the 441-residue human Tau protein. The approach complements conventional multidimensional or residue type-selective assignment strategies by providing an efficient means of traversing assignment interruptions and rapidly characterizing functionally important segments in intrinsically disordered proteins.

Biomolecular NMR spectroscopy provides site-resolved information on protein structure, dynamics, interactions, and chemical modifications under near-physiological conditions. This makes NMR particularly valuable for problems in biochemistry and structural biology where conformational heterogeneity, weak interactions, transient states, or intrinsic disorder are central to function and difficult to capture by static structural techniques. Intrinsically disordered proteins and regions (IDPs/IDRs) are prominent examples: their biological activity is often regulated by local sequence motifs, post-translational modifications, and short-lived binding events, all of which can be followed by NMR with atomic-level resolution (Botova et al. 2024; Jensen 2025; Oldfield and Dunker 2014; Theillet et al. 2012; Bugge et al. 2025; He et al. 2024; Martinez-Yamout et al. 2023; Berlow et al. 2022).

A persistent practical limitation of biomolecular NMR is resonance assignment, which links NMR signals to individual atoms in the molecular structure and is essential for nearly all quantitative analyses. The conventional approach relies on multidimensional triple-resonance experiments (Cavanagh et al. 2007; Narayanan et al. 2010) followed by extensive manual or semi-automatic analysis to achieve a complete backbone assignment. While this strategy is generally effective for small to medium-sized folded proteins, it becomes increasingly challenging for large proteins and particularly for IDPs/IDRs because of severe spectral overlap, limited chemical-shift dispersion, data sparsity, repetitive low-complexity sequences, and rapid exchange of amide protons with water at neutral pH (Croke et al. 2008; Dempsey 2001).

As a result, even well-studied proteins may contain assignment gaps that coincide with functionally important regions. This creates a mismatch between experimental effort and biological need. Many studies require reliable assignment of only a limited region surrounding a modification, interaction site, or other functional hotspot region rather than the entire protein. Nevertheless, conventional experiments are optimized for uniform broadband performance, requiring substantial acquisition and analysis efforts while often failing in precisely the crowded or exchange-broadened regions of greatest interest. These limitations are particularly pronounced for large IDPs, where residue-specific differences in relaxation, solvent exchange rates, sequence context, and spectral overlap strongly affect assignment success.

We introduced FOcused SpectroscopY (FOSY) as a strategy for the local de novo assignment of biomolecular hotspots (Lesovoy et al. 2021). Rather than pursuing a complete backbone assignment, FOSY selectively targets an individual spin system using known resonance frequencies. A series of complementary selective two-dimensional experiments then enables a local sequential walk-through of neighbouring residues. By focusing the analysis to the region of interest, the method avoids much of the experimental and analysis burden associated with global assignment.

In its original demonstration (Lesovoy et al. 2021), FOSY enabled the identification of phosphorylation sites introduced by glycogen synthase kinase 3β (GSK3β) in the full-length 441-residue human Tau isoform (hTau441). Using local sequence information obtained from a small number of selective two-dimensional experiments, the method unambiguously assigned the phosphorylated residues S404 and S409.

The successful implementation of the first FOSY experiments demonstrated that targeted local assignment is feasible even in spectroscopically challenging biomolecular systems. Beyond establishing this proof of principle, it opened the way to extending the FOSY concept into a broader class of protein NMR experiments. The focused acquisition strategy provides a flexible framework for designing site-tailored experiments that optimize polarization-transfer pathways, spectral resolution, sensitivity, and measurement time, thereby broadening the applicability of FOSY to an increasingly diverse range of structural and functional studies.

A defining feature that distinguishes FOSY from conventional multidimensional NMR experiments is the use of frequency-selective polarization-transfer (SPT) elements. By fixing the frequencies of previously assigned nuclei, SPT elements eliminate unnecessary indirect dimensions while enabling highly efficient polarization transfer along the desired spin pathway. Here, to design new experiments, we introduce two new SPT modules (**Figure S1**): three-spin longitudinal single field selective polarization transfer (3S-LSF), and three-spin single field polarization transfer (3S-SFPT). These and previously published SPT schemes (Freeman 1991; Kupce and Freeman 1993; Pelupessy and Chiarparin 2000; Khaneja et al. 2003; Lesovoy et al. 2021; Korzhnev et al. 2005) are combined according to the local spin topology and the required transfer steps. Consequently, FOSY retains the spectral dispersion of 4D to 7D experiments while yielding sensitive and highly simplified 2D spectra for acquisition and interpretation.

We revise and extend the original FOSY workflow by introducing a new assignment strategy of walking forwards along the amino acid sequence from the amide group of residue (i) to the next residue (i+1). To further increase the flexibility and applicability of the FOSY toolbox to IDPs and IDRs, we complemented the original experiments employing TROSY-type relaxation optimization (Pervushin et al. 1997) with a new set of experiments specifically designed to exploit fast selective longitudinal relaxation (Pervushin et al. 2002; Schanda and Brutscher 2005) and, in addition, to tolerate rapid exchange of amide protons with water.

The experiment set illustrated in **Figure 1** comprises revised versions of the FOSY experiments hnco(CA)NH and hncoCA(N)H for the backward-walk (**Figure 1a, S2a,c**), together with two novel 2D forward-walk experiments, hnca(CO)NH and hncaCO(N)H (**Figure 1b, S2b,d**). The latter pair implements an effective 7 and 6D transfer scheme that propagates assignment information from $H_{i-1}$, $N_{i-1}$, and $C\alpha_{i-1}$ to $H_i$, $N_i$, and $CO_{i-1}$, thereby providing a directional complement to the original backward-walk strategy.

In the experiment names, a lowercase letter denotes a nucleus whose resonance frequency is fixed by selective polarization transfer (SPT) step, whereas a capital letter denotes a resonance that evolves during one of the chemical-shift evolution periods. Nuclei shown in parentheses are neither frequency-fixed nor evolved; they are included solely to indicate the magnetization-transfer pathway.

To complement the two sequential-walk strategies, we also introduce two new out-and-back experiments that determine the carbonyl and Cα chemical shifts associated with a selected amide-centered spin system. The amide proton detected hncaCO experiment (**Figure 1c, S3a**) determines $CO_{i-1}$ for known $H_i$, $N_i$, and $C\alpha_i$ frequencies, whereas the hncoCA experiment (**Figure 1d, S3b**) determines $C\alpha_i$ for known $H_i$, $N_i$, and $CO_{i-1}$ frequencies. For a selected spin system, these experiments provide verification that the detected $CO_{i-1}$, $C\alpha_i$, $N_i$, and $H_i$ resonances belong to the same spin system and therefore provide information similar to a very-high-resolution 4D $H_iN_iCA_iCO_{i-1}$ experiment. An optional extension of hncaCO experiment further incorporates $C\beta_i$ editing, providing residue subtype information through direct link of $C\beta_i$ to the four resonances (**Figure 1c**).

To demonstrate the applicability of Focused Spectroscopy (FOSY), we targeted sequence-specific assignment in the 441-residue human hTau40 protein focusing on the P301/G–K/P312 segment, which remains only partially assigned in the BMRB 50701

The forward- and backward-walking strategies are illustrated in **Figure 1e,f** for the assignment link between residues V309 and T310. The backward-walk begins with the out-and-back hncaCO experiment (**Figure 1c, S3a**), in which the $H_i$, $N_i$, and $C\alpha_i$ frequencies are fixed. These frequencies may, for example, be available from an incomplete conventional backbone assignment. The experiment establishes the correlation to the $CO_{i-1}$ resonance. It can additionally be repeated with selective $C\beta_i$ decoupling to probe the $C\beta_i$ chemical shift and thereby verify the residue type (**Figure 1f**).

The established $H_i$, $N_i$, and $CO_{i-1}$ frequencies are then used in the hnco(CA)NH and hncoCA(N)H experiments to advance to the preceding residue and determine $H_{i-1}$, $N_{i-1}$, and $C\alpha_{i-1}$, which provide the starting frequencies for the next backward-walk step.

Analogously, the forward-walk begins with the out-and-back hncoCA experiment, which determines $C\alpha_i$ for known $H_i$, $N_i$, and $CO_{i-1}$ frequencies, which are fixed.

The hnca(CO)NH and hncaCO(N)H experiments are then used to advance to the following residue and determine $H_{i+1}$, $N_{i+!}$, and $CO_i$, which are required for the next forward-walk step.

A single step of the assignment-walk in either direction requires two inter-residue and one out-and-back 2D experiments, optionally supplemented by one or several additional 2D experiments for residue-type identification. At approximately 10 min per 2D experiment, the total acquisition time per one step is about 1 h. Detailed pulse programs, transfer pathways, and acquisition parameters are provided in the Supporting Information.

The forward- and backward-walking strategies are complementary in two respects. First, they increase the number of possible starting points for an assignment walk, allowing it to begin either before or after the segment to be assigned. One can start from a residue with a known assignment or exploit residue-type-selective combinatorial labeling strategies (Solyom et al. 2013; Schubert et al. 2005; Löhr et al. 2015). Second, they provide a convenient means of cross-validating assignments obtained independently by walking in either direction.

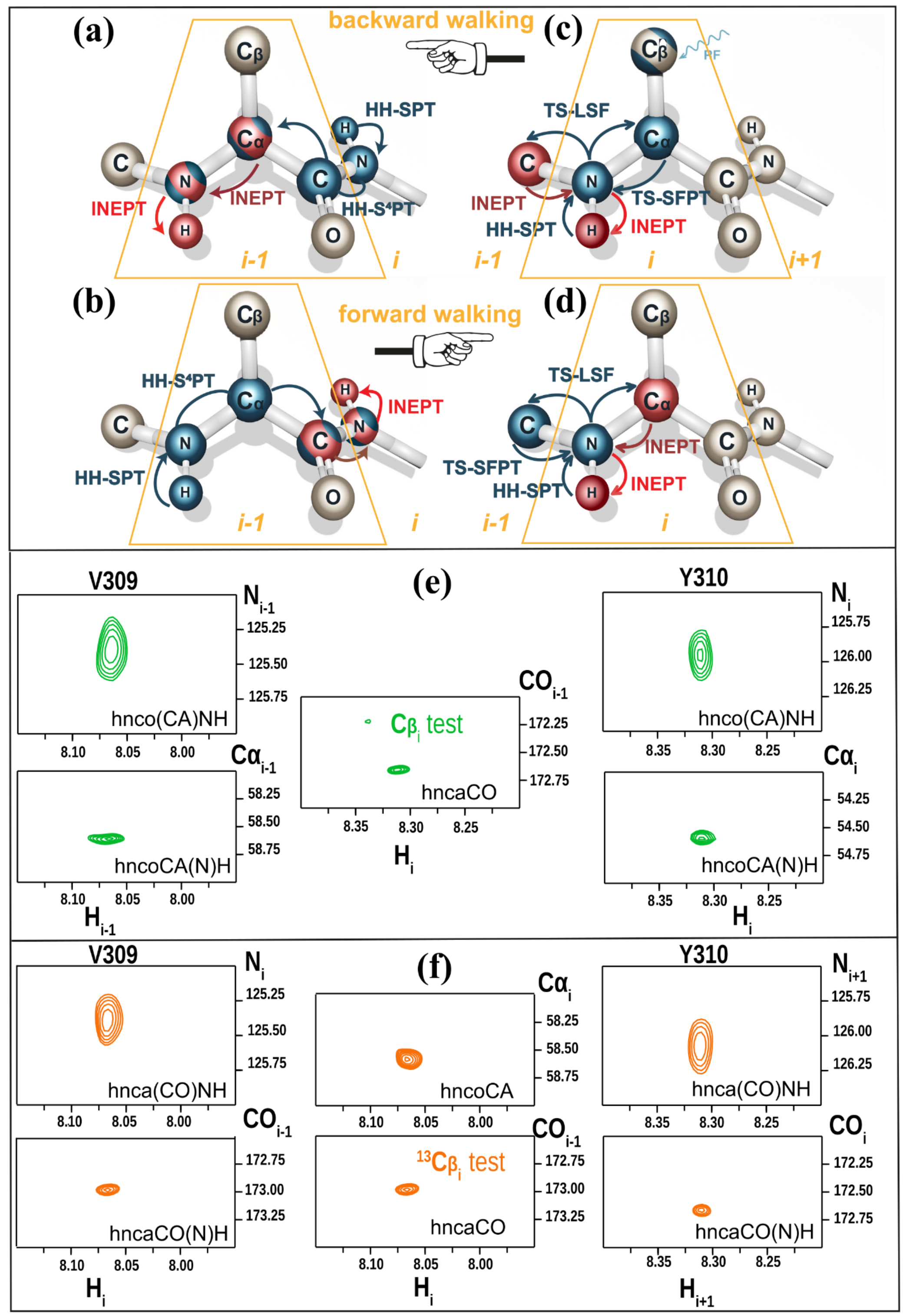

**Figure 1. Self-consistent FOSY framework for stepwise backbone assignment.** Top Panel: Schematic representation of stepwise backbone assignment using backward (a,c) and forward (b,d) FOSY experiments. Atom colour code: blue, known and fixed resonances; red, evolved in an indirect spectral dimension; striped red/blue, optionally evolved; striped grey/blue, optionally selected; grey, not used. Arrows indicate polarization transfer: blue, frequency-selective (HH-SPT, HH-S⁴PT, 3S-LSF, 3S-SFPT for description Supplementary S1); red, broadband (INEPT). Bottom Panel: example of real-time stepwise assignment shown in zoomed 2D spectra on 441-residue human hTau40 protein between two amino acids V309 and Y310 using backward (e) and forward (f) FOSY experiments.

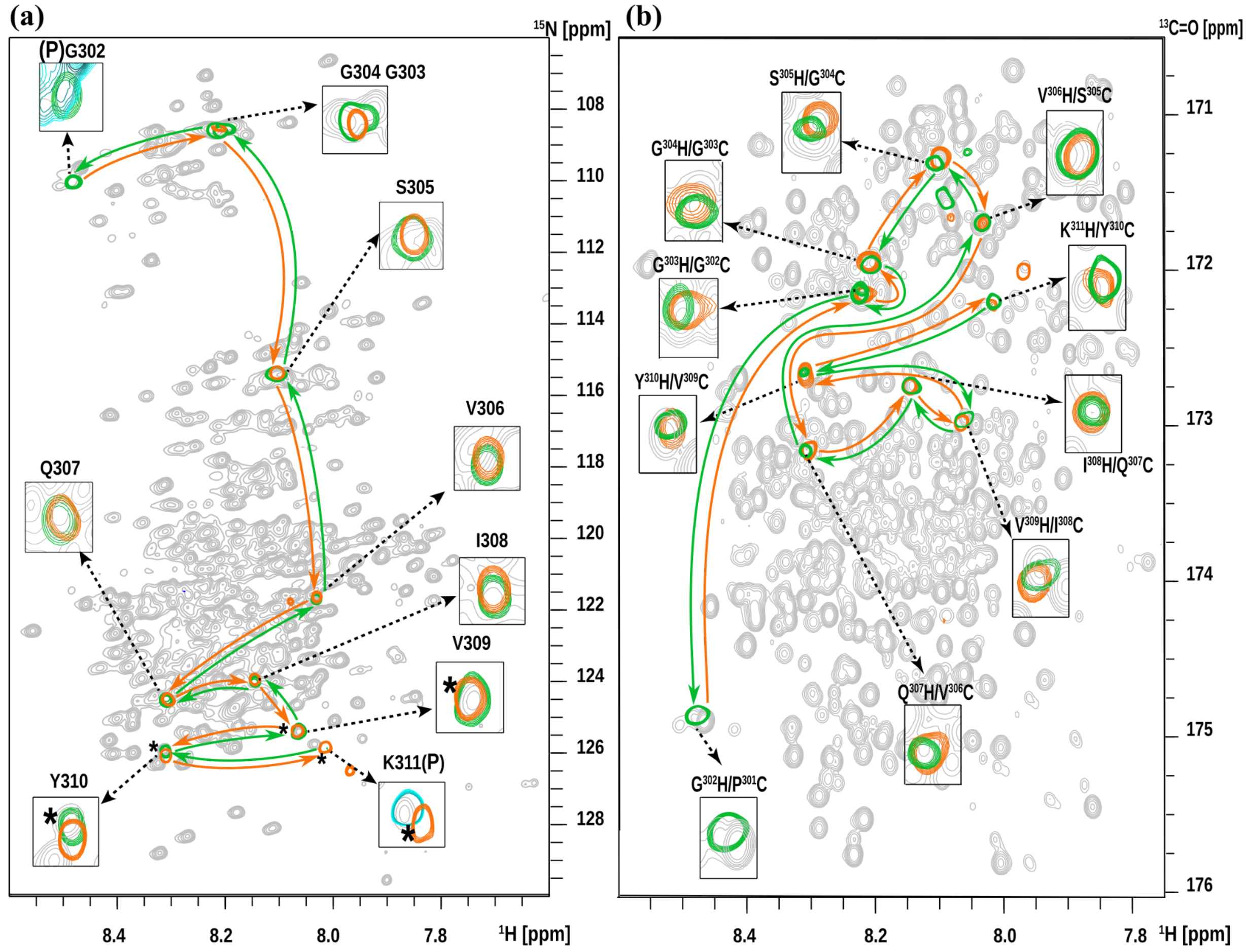

**Figure 2**. **The FOSY assignment walks in hTau40 fragment PG$^{302}$GGSVQIVYK$^{311}$P.** Backward, residues i → i−1 and forward, i → i+1 assignment walk are shown by green and orange arrows, respectively, in **(a)** $^{1}$H-$^{15}$N and **(b)** $^{1}$H-$^{13}$CO correlation spectra plotted with grey contours. Insets show zoomed superpositions of cross-peak coloured in by experiment type (a) (orange) - *hnca(CO)NH,* (green) - *hnco(CA)NH,* and (light blue) - proline-selective experiments (Lesovoy et al. 2021); (b) (orange) - *hncaCO(N)H,* (green) - *hncaCO.* In (a) asterisk indicated the peak positions of V309, Y310, and K311 corresponding to the published assignment (BMRB 50701).

From the only three consecutively assigned residues in the PG$^{302}$GGSVQ**IVYK**$^{311}$P segment, lysine K311 was the most convenient point to anchor our assignment walk, since its peak is easily identified in 2D $^{1}$H-$^{15}$N (**Figure 2a**) and $^{1}$H-$^{13}$CO (**Figure 2b**) correlation spectra providing initial $H_i$, $CO_{i-1}$, and $N_i$ chemical shifts to start the backward assignment walk. The starting position was also confirmed by observing the corresponding peak in the 2D experiment (inset K311(P) in **Figure 2a**) that shows only residues preceding prolines (−XP) (Lesovoy et al. 2021; Solyom et al. 2013). Green arrows and peak contours in **Figures 2a,b** illustrate the non-problematic backward assignment using 10–20 min 2D hnco(CA)NH, hncoCA(N)H, and hncaCO experiments. Notably, the cross peaks of V309 and Y310 assigned using the FOSY approach are in agreement with those reported previously in BMRB 50701 as indicated by black asterisks in **Figure 2a**. At each step, the residue type was confirmed by selective Cβ decoupling or by using the glycine-specific version of hncaCO(N)H experiment. The walk was concluded by assigning resonance of G302 and was confirmed by the presence of its HN signal in the proline specific (PX-) 2D experiment (inset (P)G302 in **Figure 2a**). Assignment for each residue in the stretch, was additionally confirmed by the forward steps using 2D hnca(CO)NH, hncaCO(N)H, and hncoCA (orange arrows and peak contour in **Figures 2a**), thereby illustrating the forward walk for this region of the protein.

The selective FOSY strategy presented here offers a practical solution for overcoming assignment bottlenecks in intrinsically disordered proteins and protein regions. Rather than pursuing complete backbone assignment of the entire protein, the methodology focuses experimental effort on interrupted segments, where conventional approaches fail because of severe resonance overlap, rapid amide proton exchange with water, or the absence of detectable sequential correlations. The self-consistent suite of 2D FOSY experiments presented here enables forward and backward local backbone assignment by providing unambiguous high-dimensional sequential connectivity in both directions across the most challenging and functionally important regions of the protein sequence, while preserving sensitivity and avoiding the time- and labour-intensive acquisition and analysis of high-dimensional spectra. Rather than replacing conventional multidimensional(Cavanagh et al. 2007) and residue type-selective(Schubert et al. 2005; Agback et al. 2020) assignment strategies, FOSY complements them by providing an efficient means of traversing assignment interruptions and rapidly characterizing functional hotspots in intrinsically disordered regions.

## Acknowledgements

The authors thank the Swedish NMR Centre for access to the instruments and support. This work was supported by Swedish Research Council 2023-03485 to V.O. This study used NMRbox: National Center for Biomolecular NMRData Processing and Analysis, a Biomedical Technology Research Resource (BTRR), which is supported by NIH grant P41GM111135 (NIGMS).

## Referencess

**Supporting Information**

# FOSY: Segmental Backbone Assignment in Intrinsically Disordered Proteins

*Dmitry M. Lesovoy*, Tatiana Agback , Panagiota S. Georgoulia, Irena Matečko-Burmann, Björn M. Burmann, and Vladislav Y. Orekhov**

### Protein expression and purification

Expression and purification of full-length hTau40, including the production of uniformly $^{2}H,^{15}N,^{13}C$-labelled protein, were described in detail previously (Lesovoy et al. 2021). Briefly, hTau40 was expressed in *E. coli* BL21(λDE3) Star™ cells from a modified pET28b construct encoding hTau40 N-terminally fused to His-SUMO-tag. The isotopically labelled protein was produced in M9 minimal medium using $^{15}NH_4Cl$ and D-$^{2}H^{13}C$-labelled glucose, with $^{2}H,^{15}N,^{13}C$-labelled protein produced in $^{2}H_2O$-based medium. Following induction with isopropyl-thiogalactoside (IPTG), cells were harvested and lysed, and hTau40 was purified by His-affinity chromatography. The His-SUMO tag was removed by SenP1 proteolytic cleavage, followed by a second His-affinity purification and final size-exclusion chromatography in NMR buffer (25 mM sodium phosphate, pH 6.9, 50 mM NaCl, 1 mM EDTA). Purified hTau40 was concentrated to approximately 500 μM, flash-frozen in liquid nitrogen, and stored at −80 °C until use. For full experimental details, see our previous publication (Lesovoy et al. 2021).

### Experimental NMR technical details

All NMR experiments were recorded at 15 °C on an 900 MHz Bruker AVANCE IIIHD spectrometer equipped with a 3 mm TCI $^{1}H,^{15}N,^{13}C$ cryoprobe. Samples contained 150 μM uniformly $^{2}H,^{15}N,^{13}C$ -labelled hTau40 dissolved in the NMR buffer (25 mM sodium phosphate buffer pH 6.9, 50 mM NaCl, 1 mM EDTA) prepared in 90% $^{1}H_2O$/10% $^{2}H_2O$ 9:1). Before use, the samples were stored at −80 °C.

Spectra were acquired, processed, and analysed using TopSpin 3.5pl7 (Bruker BioSpin). Acquisition parameters for the 2D hnco(CA)NH, hncoCA(N)H, hnca(CO)NH, hncaCO(N)H, hncaCO and hncoCA pulse sequences are provided in **Table S1** and in the legends of **Figures S2** and **S3**.

### Selective Polarization Transfer (SPT) blocks

The newly designed in this work $^{15}N$-selective 3S-LSF block shown in (**Figure S1a**) selectively transfers Nz magnetization to NzCAzCOz with selectivity on $\boldsymbol{\nu}_N$. The $^{13}C$-selective 3S-SFPT blocks are shown in (**Figure S1b,c**). The block in (b) provides selective transfers -NyCAzCOz to -Ny with selectivity on $\boldsymbol{\nu}_{CO}$, whereas the block in (c) provides selective transfer -NyCAz(CB)COz to -Ny with selectivity on $\boldsymbol{\nu}_{CA}$ and $\boldsymbol{\nu}_{CB}$.

In the schematic representation of the irradiation frequencies for the 3S-LSF block in **Figure S1a**, the continuous-wave (cw) pulse, denoted Ncw, consists of a single frequency $^{15}N$ continuous wave irradiation. The B1 field strength of the Ncw irradiations (Ncw1, Ncw2, Ncw3, Ncw4 and Ncw5) used in individual experiments (**Figures S2 and S3)** are listed in **Table S1** in Hz units.

For the 3S-SFPT block in **Figure S1b**, the Ccw is a sandwich consisting of two continuous-waves irradiations at $\boldsymbol{\nu}_{CO\,i-1} \pm {}^{1}\boldsymbol{J}_{CACO}/2$ frequencies. The irradiation parameters for individual Ccw4, Ccw5, and Ccw6 are listed in **Table S1**.

For the 3S-SFPT block in **Figure S1c,** the CAcw sandwich is a four-continuous-wave scheme, with a version marked by an asterisk for optional $C\beta_i$ decoupling and "G" for glycines. Parameters for Ccw1*, Ccw2*, Ccw3* and Ccw1, Ccw2, Ccw3 as well as glycine versions are found in **Table S1.**

Without the Cβi decoupling, the four continuous-waves irradiations are applied at $\boldsymbol{\nu}_{\mathbf{CA}i} \pm {}^1\boldsymbol{J}_{CACO}/2 \pm {}^1\boldsymbol{J}_{CACB}/2$ frequencies with the same B1 field strength.

In the version with the $C\beta_i$ decoupling as well as for glycines, the cw sandwiches comprise four continuous-wave irradiations. While the two applied at $\boldsymbol{\nu}_{CAi} \pm {}^1\boldsymbol{J}_{CACO}/2$ are used for the polarisation transfer, the irradiation at $\boldsymbol{\nu}_{CBi}$, serves for the concurrent $C\beta_i$ decoupling and the one given at $2\boldsymbol{\nu}_{CAi}$-$\boldsymbol{\nu}_{CBi}$ compensates the off-resonance effect of the decoupling cw on $C\alpha_i$. A moderate B1 decoupling field strength (e.g. B1≈300Hz) is used to provide a narrow band selective coverage for the amino acid type discrimination.

All three transfer blocks were designed to preserve Nz magnetization and the inphase $^{15}N$ coherence (Nx, Ny) with respect to the amide proton during long $^1J_{NCO}$ and $^1J_{NCA}$ evolution periods, thereby minimizing losses due to proton-water exchange. In particular, selective $^1H$ decoupling is used to suppress $^1J_{NH}$ evolution and prevent buildup of NzHz and NxyHz terms.

The simulated spin system evolution within the 3S-LSF and 3S-SFPT blocks are shown in **Figure S1d** and **Figure S1e,f** respectively. The magnetization transfer efficiency approach 100%, if only the active couplings are considered in the simulations (**Figure S1g-i**), and exceeding 50% at realistic relaxation rates ($^{13}CO$ $R_2$=20Hz, $^{15}N$ $R_2$=10Hz, and $^{13}C\alpha$ $R_2$=2Hz) and the passive scalar coupling $^2J_{NCA}$=5Hz is plotted in **Figure S1j-l** versus the magnetization transfer time, and the B1 field strength.

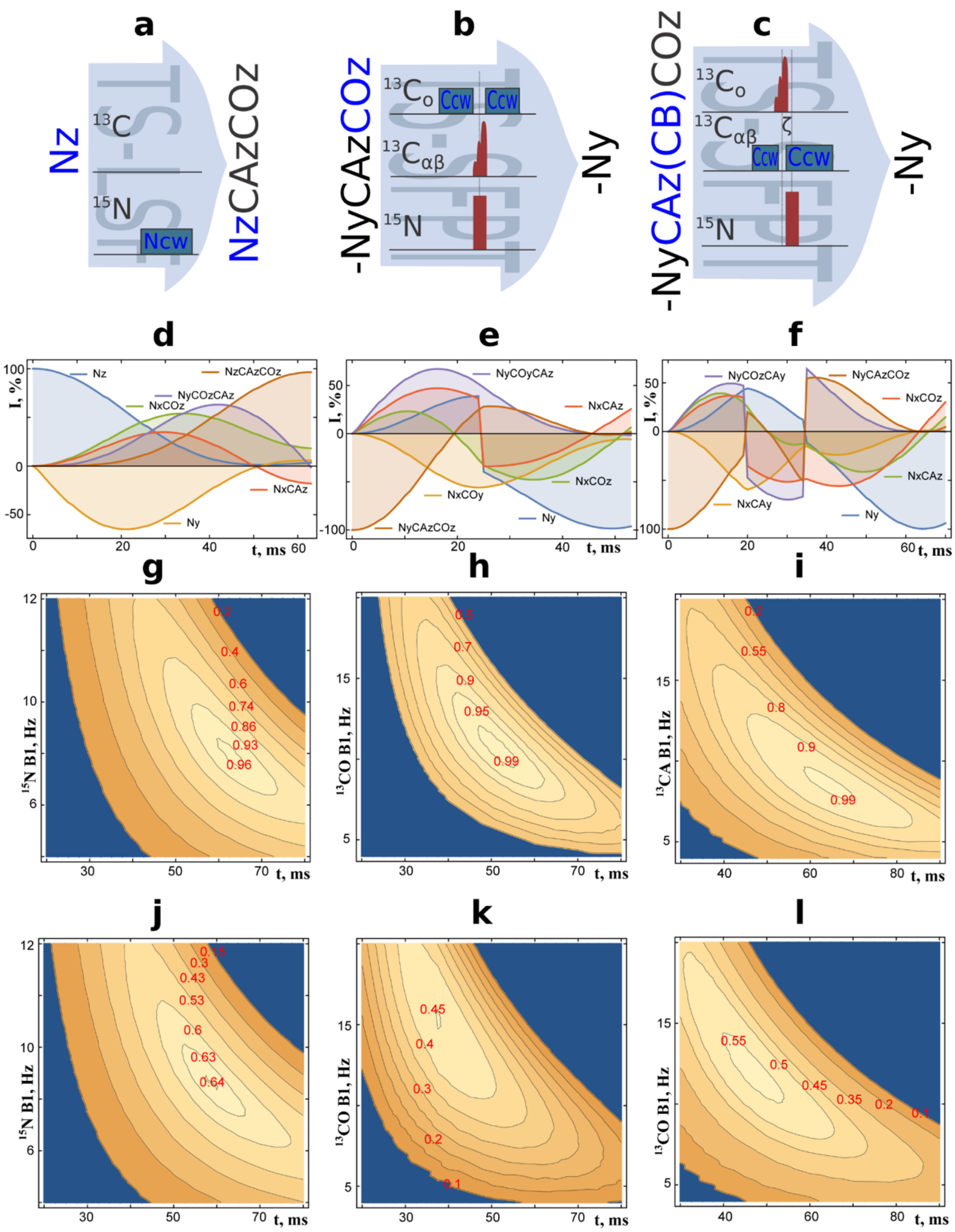

**Figure S1**. (a) The originally designed $^{15}$N-selective 3S-LSF block; (b,c) The 3S-SFPT $^{13}$C-selective blocks where the 3S-SFPT block in (b) transfers the coherence selectively with respect to $\boldsymbol{\nu}_{CO}$, whereas the block in (c) transfers coherence selectively with respect to $\boldsymbol{\nu}_{CA}$ and $\boldsymbol{\nu}_{CB}$. The coherence evolution within the 3S-LSF block (d) and the 3S-SFPT blocks is shown in (e) for $\boldsymbol{\nu}_{CO}$ selective transfer and in (f) for $\boldsymbol{\nu}_{CA}/\boldsymbol{\nu}_{CB}$ selective transfer. The magnetization transfer efficiencies (ranging from 0 to 1) are simulated and presented in panels (g-l) as a function of the corresponding block length, t (ms) and the heteronuclear B1 field strength (Hz) of the corresponding continuous-wave (cw) irradiation applied to $^{15}$N or $^{13}$C in panels (a-c). Particularly for panels (g-i) the simulations were performed considering only the active scalar couplings $^{1}J_{NCO}$=15Hz and $^{1}J_{NCA}$= 12Hz. In panels (j-l) the transverse relaxation rates $^{13}$CO $R_2$=20Hz, $^{15}$N $R_2$=10Hz, $^{13}$Cα $R_2$=2Hz, together with the passive scalar coupling $^{2}J_{NCA}$= 5Hz were taken into account.

**Table S1** The parameters of the continuous wave (cw) blocks used in the SPT modules of the FOSY pulse sequences presented in **Figures S2** and **S3**. The versions of the cw blocks marked with asterisks include irradiation at frequency of the $C\beta_i$ carbon for homodecoupling of $C\alpha_i$ with respect to $^{1}J_{CACB}$. The versions of the cw blocks marked with "G" were used for glycines.

| Cw nucleus | Frequencies | Phase | Phase alignment | B1 field strength, Hz | Duration ms |
|---|---|---|---|---|---|
| Hcw1 | $\nu_{Hi}$ | x | start | 99 | 10 |
| Hcw2 | $\nu_{Hi}$ | x | middle | 500-700 | decoupling |
| Hcw3 | $\nu_{Hi}$ | x | start | 42 | 10 |
| Ncw1 | $\nu_{Ni}$ | x | end | 99 | 10 |
| Ncw2 | $\nu_{Ni}$ | x | middle | 8.8 | 55 |
| Ncw3 | $\nu_{Ni}$ | x | end | 42 | 10 |
| Ncw4 | $\nu_{Ni}$ | x | start | 16 | 55 |
| Ncw5 | $\nu_{Ni}$ | x | start | 14 | 63 |
| Ccw1 | $\nu_{CAi} - {}^1J_{CACO}/2 - {}^1J_{CACB}/2$<br>$\nu_{CAi} - {}^1J_{CACO}/2 + {}^1J_{CACB}/2$<br>$\nu_{CAi} + {}^1J_{CACO}/2 - {}^1J_{CACB}/2$<br>$\nu_{CAi} + {}^1J_{CACO}/2 + {}^1J_{CACB}/2$ | x<br>x<br>x<br>x | end | 9.5<br>9.5<br>9.5<br>9.5 | 16.7 |
| Ccw2 | $\nu_{CAi} - {}^1J_{CACO}/2 - {}^1J_{CACB}/2$<br>$\nu_{CAi} - {}^1J_{CACO}/2 + {}^1J_{CACB}/2$<br>$\nu_{CAi} + {}^1J_{CACO}/2 - {}^1J_{CACB}/2$<br>$\nu_{CAi} + {}^1J_{CACO}/2 + {}^1J_{CACB}/2$ | x<br>x<br>x<br>x | start | 9.5<br>9.5<br>9.5<br>9.5 | 32.7 |
| Ccw3 | $\nu_{CAi} - {}^1J_{CACO}/2 - {}^1J_{CACB}/2$<br>$\nu_{CAi} - {}^1J_{CACO}/2 + {}^1J_{CACB}/2$<br>$\nu_{CAi} + {}^1J_{CACO}/2 - {}^1J_{CACB}/2$<br>$\nu_{CAi} + {}^1J_{CACO}/2 + {}^1J_{CACB}/2$ | x<br>x<br>-x<br>-x | end | 14<br>14<br>14<br>14 | 63 |
| Ccw1* | $\nu_{CBi}$<br>$\nu_{CAi} - {}^1J_{CACO}/2$<br>$\nu_{CAi} + {}^1J_{CACO}/2$<br>$2\nu_{CAi} - \nu_{CBi}$ | x<br>x<br>x<br>-x | end | 200-400<br>9.5<br>9.5<br>200-400 | 16.7 |
| Ccw2* | $\nu_{CBi}$<br>$\nu_{CAi} - {}^1J_{CACO}/2$<br>$\nu_{CAi} + {}^1J_{CACO}/2$<br>$2\nu_{CAi} - \nu_{CBi}$ | x<br>x<br>x<br>-x | start | 200-400<br>9.5<br>9.5<br>200-400 | 32.7 |
| Ccw3* | $\nu_{CBi}$<br>$\nu_{CAi} - {}^1J_{CACO}/2$<br>$\nu_{CAi} + {}^1J_{CACO}/2$<br>$2\nu_{CAi} - \nu_{CBi}$ | x<br>x<br>-x<br>-x | end | 200-400<br>14<br>14<br>200-400 | 63 |
| Ccw1$^G$ | $\nu_{CAi} - {}^1J_{CACO}/2$<br>$\nu_{CAi} + {}^1J_{CACO}/2$ | x<br>x | end | 9.5<br>9.5 | 16.7 |
| Ccw2$^G$ | $\nu_{CAi} - {}^1J_{CACO}/2$<br>$\nu_{CAi} + {}^1J_{CACO}/2$ | x<br>x | start | 9.5<br>9.5 | 32.7 |
| Ccw3$^{G}$ | $\nu_{CA}$i - $^{1}J_{CACO}/2$<br>$\nu_{CA}$i + $^{1}J_{CACO}/2$ | x<br>-x | <br>end | 14<br>14 | 63 |
| Ccw4 | $\nu_{CO}$i-1 - $^{1}J_{CACO}/2$<br>$\nu_{CO}$i-1 + $^{1}J_{CACO}/2$ | x<br>x | end | 14<br>14 | <br>20 |
| Ccw5 | $\nu_{CO}$i-1 - $^{1}J_{CACO}/2$<br>$\nu_{CO}$i-1 + $^{1}J_{CACO}/2$ | x<br>x | start | 14<br>14 | <br>20 |
| Ccw6 | $\nu_{CO}$i-1 - $^{1}J_{CACO}/2$<br>$\nu_{CO}$i-1 + $^{1}J_{CACO}/2$ | x<br>-x | end | 16<br>16 | <br>55 |

**Pulse sequences description**

All pulse sequences shown in **Figures 1**, **S2**, and **S3** start with a moderately selective 90° $^{1}$H pulse, followed by a heteronuclear Hartmann–Hahn selective polarization transfer (HH-SPT) step (Lesovoy et al. 2021) which selectively transfers amide $H_ix$ magnetisation to $N_ix$ (points 1-3 steps in **Figures S2**, **S3**).
The moderately selective 90° $^{1}$H pulse minimally perturbs water and other non-amide protein protons, allowing their longitudinal magnetization to recover rapidly. The subsequent HH-SPT block generates in-phase $^{15}$N transverse coherence ($N_ix$) which is unaffected by amide proton–water exchange because the magnetization resides on the $^{15}$N spin during the transfer. During most of the subsequent parts of the pulse sequence, the exchange-induced signal losses are largely avoided provided the exchange rate remains significantly smaller than the $^{1}J_{NH}$ coupling. Furthermore, selective continuous waves $^{1}$H decoupling suppresses $^{1}J_{NH}$ evolution during this period, preventing the formation of antiphase (NzHz) and single-quantum (NxyHz) coherences.

***hnco(CA)NH*** *and* ***hncoCA(N)H*** *experiments.*

The interresidual experiments ***hnco(CA)NH*** and ***hncoCA(N)H*** for backward sequential assignment (**Figures 1a** and **S2a,c**) define the chemical shifts of $C\alpha_{i-1}$, $N_{i-1}$, and $H_{i-1}$ by fixing the known $CO_{i-1}$, $N_i$, and $H_i$ resonances (**Figure 1c**). Following the HH-SPT step (10 ms simultaneous proton and nitrogen continuous-wave irradiation, Hcw3 and Ncw3, respectively, with B1 = 42 Hz; **Table S1**), a Hartmann–Hahn spin-state selective polarization transfer (HH-S$^{4}$PT) (Lesovoy et al. 2021) module transfers in-phase $N_ix$ coherence to $CO_{i-1}xCA_{i-1}z$ (point 4 in **Figure S2a,c**). HH-S$^{4}$PT is the selective 55ms-long irradiation at $N_i$ and $CO_{i-1}$ frequencies (Ncw4 and Ccw6, respectively; B1=16Hz; **Table S1**) with simultaneous selective $^{1}$H decoupling (Hcw2, B1≈ 500 Hz; **Table S1**). The overall frequency selectivity of the experiments is defined by the B1 strength used in the HH-SPT and HH-S$^{4}$PT blocks and is approximately ~42, 16, and 16 Hz for $H_i$, $N_i$, and $CO_{i-1}$, respectively.
The subsequent INEPT-type transfer converts $CO_{i-1}zCA_{i-1}y$ (point 6 **Figures S2a,c**) coherence into $CA_{i-1}yN_{i-1}z$ (8 point in **Figure S2a,c**) under evolution over the active couplings $^{1}J_{COi-1CAi-1}$ and $^{1}J_{CAi-1Ni-1}$.
The $N_{i-1}$ constant-time chemical shift evolution in ***hnco(CA)NH*** is implemented between points 9 and 10 (**Figures S2a**), whereas the $C\alpha_{i-1}$ constant-time chemical shift evolution in ***hncoCA(N)H*** is introduced between points 6 and 7 (**Figures S2c),** yielding a 2D $C\alpha_{i-1}/H_{i-1}$ spectrum. Amide region-selective $^{1}$H decoupling using **p5m4sp** block between points 9 and 10 in both pulse sequences maintains the nitrogen coherence in phase with respect to the proton, thereby preventing signal losses due to amide proton-water exchange.
At the end of both pulse sequences, between points 10 and 11, a conventional sensitivity enhanced (a) or refocused (c) INEPT step is used to transfer the $N_{i-1}xH_{i-1}z$ anty-phase coherence “back” to in-phase $H_{i-1}x$ magnetisation.
In summary, the following main magnetization terms are created in the course of the experiments at the marked points (**Figure S2a,c):** (1) $H_iz$, (2) $H_ix$, (3) $N_ix$, (4) $CO_{i-1}xCA_{i-1}z$ , (5) $CO_{i-1}zCA_{i-1}z$, (6) $CO_{i-1}zCA_{i-1}y$, (7) $CA_{i-1}yN_{i-1}z$, (8) $CA_{i-1}zN_{i-1}z$, (9) $CA_{i-1}zN_{i-1}x$, (10) $N_{i-1}xH_{i-1}z$, (11) $H_{i-1}x$.

***hnca(CO)NH and hncaCO(N)H*** *experiments.*

Two 2D experiments ***hnca(CO)NH*** and ***hncaCO(N)H,*** schematically shown in **Figures 1b,d,f,** are introduced in this work for the “forward walking” strategy that establishes sequential connectivity from residue *i-1* to residue *i.*

The known frequencies of $H_{i-1}$, $N_{i-1}$, and $C\alpha_{i-1}$ resonances are used to define $CO_{i-1}$, $H_i$, $N_i$ chemical shifts in the complementary 2D $CO_{i-1}/H_i$ and $N_i/H_i$ correlation experiments (**Figure S2b,d**).

At point 3 in both pulse sequences (**Figure S2b,d**), in-phase $N_{i-1}x$, coherence is established after the heteronuclear Hartmann–Hahn selective polarization transfer (HH-SPT) module. Then, the selective heteronuclear Hartmann–Hahn spin-state selective polarization transfer (HH-S$^4$PT) produces $CA_{i-1}xCO_{i-1}z$ coherence.

The effective selectivity, defined by the B1 strength in the HH-SPT and HH-S$^4$PT blocks (**Table S1**), for $H_{i-1}$, $N_{i-1}$, and $C\alpha_{i-1}$ frequencies are approximately 42, 14, and 14 Hz, respectively. Optionally, selective $C\beta_{i-1}$ decoupling (Ccw3* B1 ≈ 200–700 Hz) can be applied to suppress $^1J_{CAi-1CBi-1}$ scalar coupling splitting, thereby improving both selectivity and sensitivity of the experiments.

The remaining parts of the experiments (**Figures S2b,d**) are similar to the corresponding blocks in the backward ***hnco(CA)NH*** and ***hncoCA(N)H*** experiments (**Figures S2a,c**) and include chemical-shift evolution and INEPT-based transfer blocks. In summary, the following main magnetization terms are created during the experiments at the marked points (**Figure S2b,d)**: (1) $H_{i-1}z$, (2) $H_{i-1}x$, (3) $N_{i-1}x$, (4) $CA_{i-1}xCO_{i-1}z$ , (5) $CA_{i-1}zCO_{i-1}z$, (6) $CA_{i-1}zCO_{i-1}y$, (7) $CO_{i-1}yN_iz$, (8) $CO_{i-1}zN_iz$, (9) $CO_{i-1}zN_ix$, (10) $N_ixH_iz$, (11) $H_ix$.

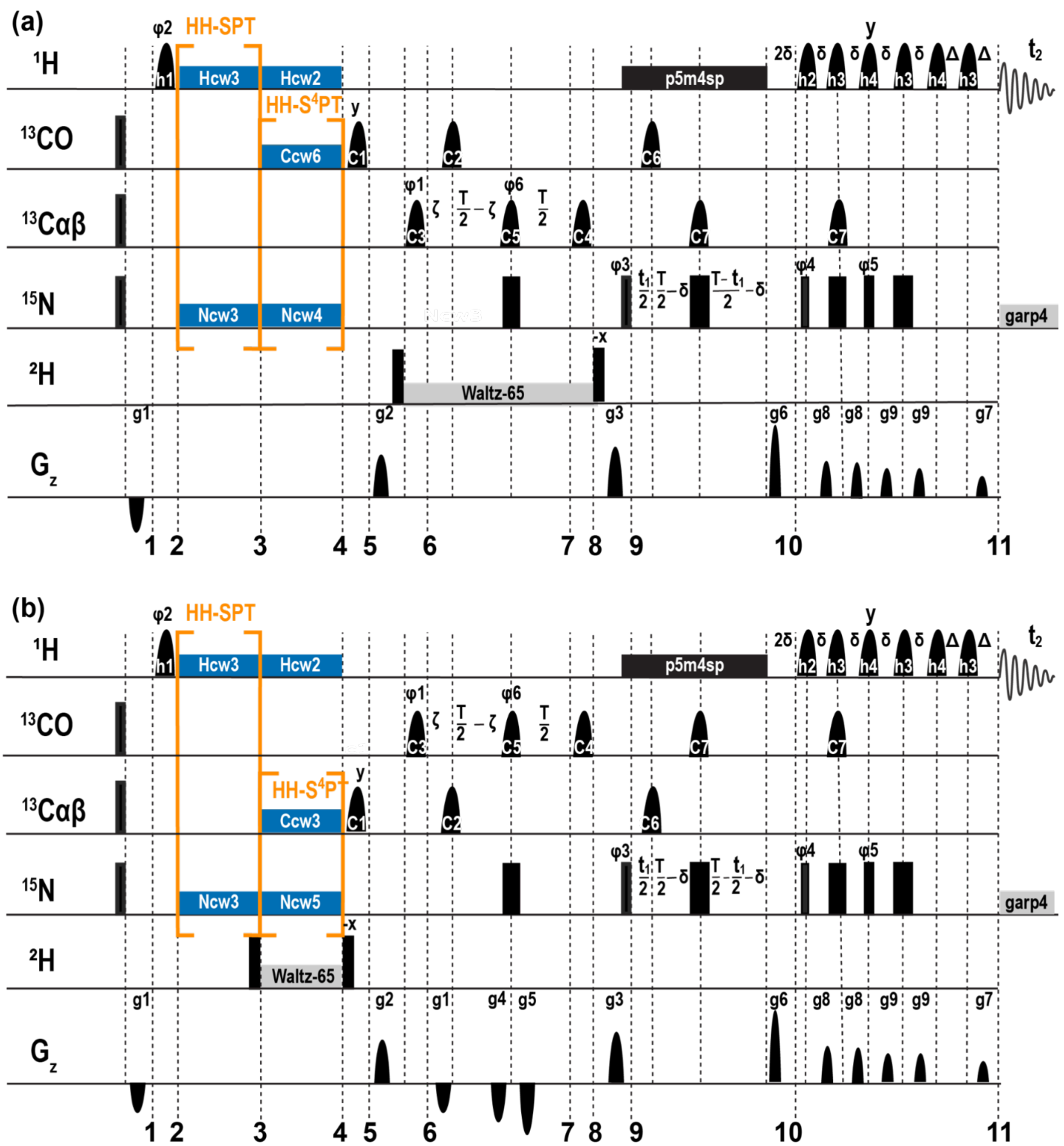

**Figure S2**. **2D inter-residual FOSY pulse programs for U-$^{15}$N/$^{13}$C/$^{2}$H labeled proteins in an $H_2O$-based buffer solution.** (a) hnco(CA)NH; (b) hnca(CO)NH; (c) hncoCA(N)H and (d) hncaCO(N)H. Blue elements indicate frequency selective elements. Rounded elements represent frequency band-selective and pulse field gradient (PFG) pulses. Narrow and thick bars represent 90° and 180° high power pulses, respectively. Echo/anti-echo type constant-time chemical shift evolutions is implemented for $^{15}$N using φ5, g6*EA and $t_1$ in (a) and (b), and for $^{13}$C using φ1, g6*EA and $t_1$ in (c) and φ1, g5*EA, g6*EA, g9*EA and $t_1$ in (d). A classical sensitivity-enhanced INEPT is used in (a) and (b) for the 2D experiments with $^{15}$N chemical shift evolution, whereas a classical refocused INEPT is used in (c) and (d) for the experiments with $^{13}$C chemical-shift evolution. The constant-time delays, T, are 13ms for (a) and (c) and 14ms for (b) and (d). The delays were $\zeta = 1/4\ ^1J_{CACO}$ = 4.7 ms and $\delta \approx 1/4\ ^1J_{NH}$ = 2.75 ms, and Δ = 0.7 ms. Selective pulses on amide $^1$H were as follows:: h1(90°) = 0.8 ms Sinc pulse at the selected $\boldsymbol{\nu}$Hi; h2(90°) = 1.74 ms time reversed EBurp2 pulse, h3(180°) = 1.85 ms ReBurp pulse, and h4(90°) = 1.74 ms EBurp2 pulse Anh4(180°) and p5m4sp (Tycko et al. 1985; Fujiwara and Nagayama 1988; Anklin and Byrd 2021) band-selective decoupling were implemented using the I3Snob pulses with the same length of approximately 1ms. The exact pulse duration was calculated by maintaining an integer number of pulses in the p5m4sp decoupling sequence, which was applied at 8.5 ppm (the center of the amide $^1$H region). Selective pulses on $^{13}$C were as follows. For the backward experiments (a) and (c): c1(90°) = 600 μs Sinc pulse and c2(180°) = 600 μs Sinc pulse at the selected $\boldsymbol{\nu}_{COi}$; c3(90°) = 500 μs Q5 and c4(90°) = 500 μs time reversed Q5 pulse at 55 ppm (the center of the Cα region); c5(180°) = 200 μs Q3 pulse at 39 ppm (the center of the Cα/ Cβ region); c6(180°) = 800 μs IBurp1 pulse at 173 ppm (the center of the CO region); c7(180°) = 600 μs IBurp1 pulse at 55 ppm (the center of the Cα region). For the forward experiments (b) and (d): c1(90°) = 800 μs Sinc pulse and c2(180°) = 800 μs Sinc pulse at the selected $\boldsymbol{\nu}_{CAi}$; c3(90°) = 500 μs Q5 pulse and c4(90°) = 500 μs time reversed Q5 pulse at 173 ppm (the center of the CO region); c5(180°) = 400 μs Q3 pulse at 173 ppm (the center of the CO region); c6(180°) = 265 μs IBurp1 pulse at 39 ppm (the center of the Cα/ Cβ region); c7(180°) = 800 μs IBurp1

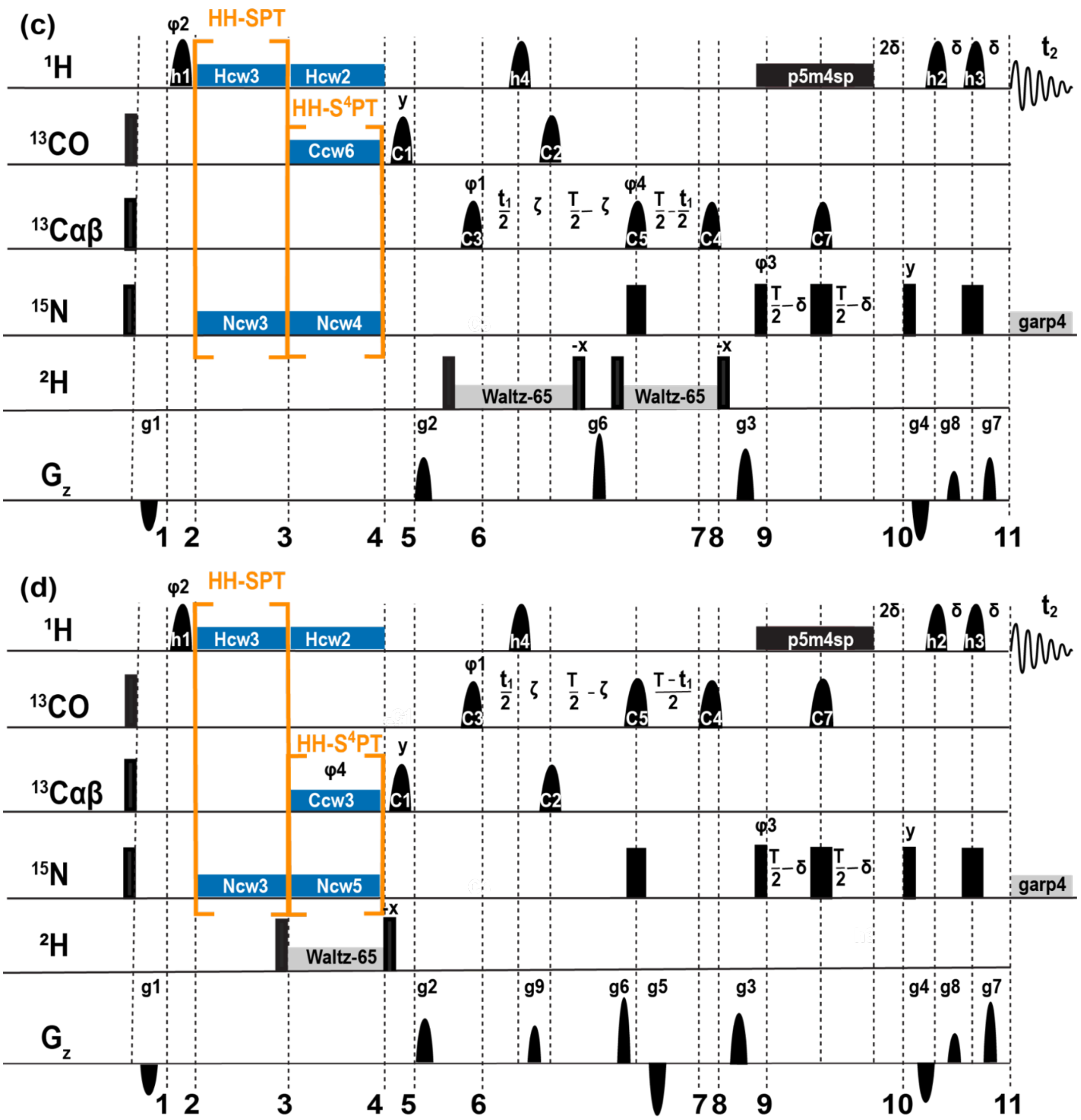

pulse at 173 ppm (the center of the CO region). The parameters of the continuous-wave blocks Hcw2, Hcw3, Ncw3, Ncw4, Ncw5,Ccw3, Ccw6 are summarized in **Table S1**. All gradient pulses have a smoothed square (SMSQ) shape. For g1-g5, the gradient-pulse duration is 1ms except for g5 in (d), which has a duration of 0.5 ms. All g6–g9 gradient pulses have a duration of 0.5 ms. The relative gradient strengths g1 = -17%, g2 = 47%, g3 = 57%, g4 = -43%, g5 = -60%, g6 = 80%, g8 = 30%, and g9 = 20% are the same for all four experiments (a- d), whereas relative gradient strength is g7 = 8.1% for 2D the $^{1}H/^{15}N$ experiments (a) and (b), and for 2D $^{1}H/^{13}C$ g7 = 50.1% for (c) and g7 = 70.2% for (d). A relative gradient strength 100% corresponds to 53.5 G/cm. The pulse phase cycles (default = x when not specified) are as follows. For the 2D $^{1}H/^{15}N$ experiments, (a): φ1=x,-x; φ2=2(-y), 2(y); φ3=4(y), 4(-y); φ4=8(x), 8(-x); φ5=8(y), 8(-y); φ6=16(x), 16(-x); and φrec=2(x, -x, -x, x, -x, x, x, -x, -x, x, x, -x, x, -x, -x, x). For (b): φ1=x,-x; φ2=2(-y), 2(y); φ3=4(y), 4(-y); φ4=8(x), 8(-x); φ5=8(y), 8(-y); φ6=16(x), 16(-x); and φrec=x, -x, -x, x, -x, x, x, -x, -x, x, x, -x, x, -x, -x, x, -x, x, x, -x, x, -x, -x, x, x, -x, -x, x, -x, x, x, -x; For the 2D $^{1}H/^{13}C$ experiments, (c): φ1=x,-x; φ2=2(-y), 2(y); φ3=4(y), 4(-y); φ4=8(x), 8(-x); and φrec=2(x, -x, -x, x, -x, x, x, -x). For (d): φ1=x,-x; φ2=2(-y), 2(y); φ3=4(y), 4(-y); φ4=8(x), 8(-x); and φrec=x, -x, -x, x, -x, x, x, -x, -x, x, x, -x, x, -x, -x, x.

***hncaCO and hncoCA*** *experiments*

The ***hncaCO*** *and* ***hncoCA*** "out-and-back" experiments are essential complementary elements in the backward (**Figure 1a,c**) and forward (**Figure 1b,d**) assignment walk strategies, respectively. Both experiments use 3S-LSF and 3S-SFPT blocks shown in **Figure S1**.

The ***hncaCO*** experiment (**Figure S3a)** is initialized using the $H_i$, $N_i$, and $C\alpha_i$ resonances, which may be obtained from the preceding inter residual backward experiment (**Figure 1a**), and returns $CO_{i-1}$ chemical shift subsequently required in the next backward step with ***hncoCA(N)H*** and ***hnco(CA)NH*** experiments (**Figure 1a**).

Similarly, ***hncoCA*** experiment (**Figure S3b)** is initialized using the $H_i$, $N_i$, and $CO_{i-1}$ resonances, to define the $C\alpha_i$ chemical shifts, thus allowing the next step forward to residue i+1 using ***hncaCO(N)H*** and ***hnca(CO)NH*** experiments.

Both experiments begin with an HH-SPT module (points 1-3), executed using 10 ms simultaneous continuous-wave irradiations at $H_i$ (Hcw1) and Ni (Ncw1) frequencies with B1 = 99 Hz (**Table S1)**.

This is followed by three 90° hard pulses applied to $^{15}N$, which transfer $N_ix$ (points 3) to $N_iz$ magnetization. It is subsequently converted between points 4-5 in **Figures S3a,b** into $N_izCA_izCO_{i-1}z$ using a newly designed $^{15}N$-selective Three Spin Longitudinal Single Field Selective Polarization Transfer (3S-LSF) module (**Figure S1a,d,j)** implemented as a 55 ms continuous-wave irradiation, at $\nu_{Ni}$, (Ncw2) with B1=8.8Hz (**Table S1**) and with simultaneous selective $^1H$ decoupling (Hcw2, B1≈ 500-700 Hz; **Table S1**), which suppresses evolution under the $^1J_{NH}$ coupling.

A 90° hard pulse applied to CO (**Figures S3a**) or Cα (**Figures S3b** ) is followed, between points 6 and 8 by conventional evolution of $CO_{i-1}$ or $C\alpha_i$, respectively. Between points 8 and 11 (**Figures 3a,b**), the Three-Spin Single Field Polarization Transfer (3S-SFPT) module shown in **Figure S1b,c** transfers $N_iyCA_izCO_{i-1}z$ coherence back to $N_izH_iz$.

In the ***hncaCO*** pulse sequence (**Figure S3a**), 3S-SFPT module uses 16.7 and 32.7 ms continuous-wave irradiation at $\nu_{CAi}$, (Ccw1 and Ccw2) with B1=9.5Hz, respectively (**Table S1**). In the ***hncoCA*** pulse sequence (**Figure S3b**), 3S-SFPT module uses 20 and 20 ms continuous-wave irradiations at $\nu_{COi-1}$, (Ccw4 and Ccw5) with B1=14Hz (**Table S1**). In both experiments the conditions are optimized to minimize relaxation losses while allowing simultaneous evolution under the active $^1J_{NCA}$ and $^1J_{NCO}$ couplings. Additionally selective $^1H$ decoupling (Hcw2, B1≈ 500-700 Hz; **Table S1**) is applied which suppresses evolution under the $^1J_{NH}$ coupling.

Optionally, in the ***hncaCO*** pulse sequence (**Figure 1c)** the module allows amino-acid-type discrimination through continuous-wave $C\beta i$ decoupling. In this case, in the pulse sequence shown in **Figure S3a**, the 3S-SFPT module uses 16.7 and 32.7 ms continuous-wave irradiation at $\nu_{CAi}$ with B1=9.5Hz, and at $\nu_{CBi}$ with B1=200-400Hz (**Table S1** Ccw1* and Ccw2*).

At the end of both pulse sequences between points 11 and 12, a selective $^1H$ pulse followed by a conventional INEPT step is used to transfer the $N_izH_iz$ coherence "back" to in-phase $H_ix$ magnetisation.

Overall frequency selectivities of the ***hncaCO*** *and* ***hncoCA*** experiments for $H_i$, $N_i$, $CO_{i-1}$, $C\alpha_i$ and optionally $C\beta_i$ frequencies are 99 Hz, 8.8Hz, 14Hz, 9.5Hz, and 200-400Hz, respectively. Relatively low selectivity for $H_i$, is inconsequential because the detection in the experiments is on the amide protons.

The following main magnetization terms are created in the course of the experiments at the marked points: **Figure S3a:** (1) $H_iz$, (2) $H_ix$, (3) $N_ix$, (4) $N_iz$, (5) $N_izCA_izCO_{i-1}z$, (6) $N_izCA_izCO_{i-1}y$, (7) $N_izCA_izCO_{i-1}xy$, (8) $N_izCA_izCO_{i-1}z$, (9) $N_iyCA_izCO_{i-1}z$, (10) $N_ixH_iz$, (11) $N_izH_iz$, (12) $H_ix$ and **Figure S3b:** (1) $H_iz$, (2) $H_ix$, (3) $N_ix$, (4) $N_iz$, (5) $N_izCA_izCO_{i-1}z$, (6) $N_izCO_{i-1}zCA_iy$, (7) $N_izCO_{i-1}zCA_ixy$, (8) $N_izCA_izCO_{i-1}z$, (9) $N_iyCA_izCO_{i-1}z$, (10) $N_ixH_iz$, (11) $N_izH_iz$, (12) $H_ix$.

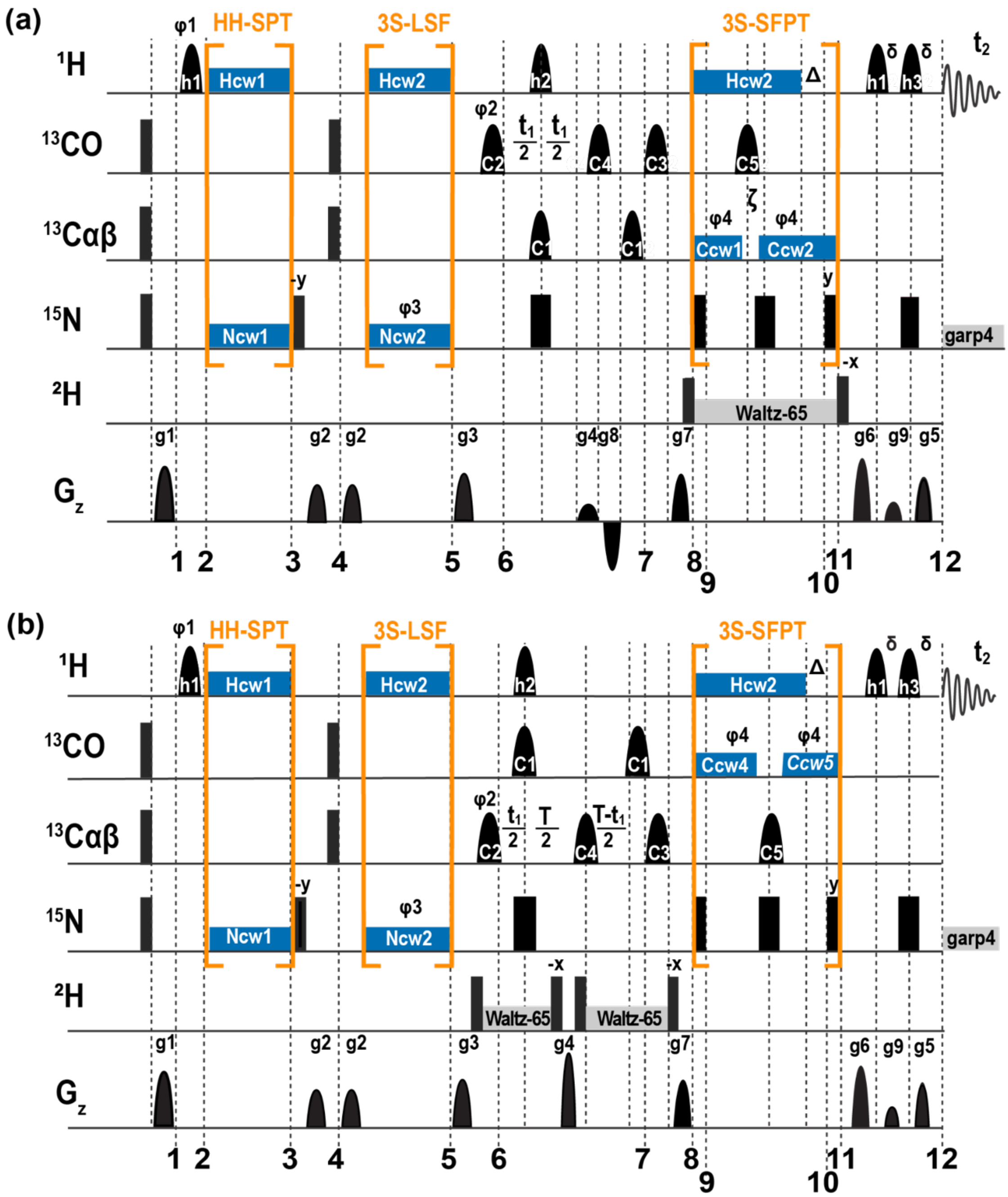

**Figure S3**. **2D annectant FOSY experiments for deuterated proteins in $H_2O$-based buffer solution.** (a) hncaCO and (b) hncoCA pulse sequences. Blue elements indicate frequency selective elements. Rounded elements represent frequency band-selective and pulse-field gradient (PFG) pulses. Echo/anti-echo type sampling of $CO_{i-1}$ using φ2, g4*EA, g8*EA, and $t_1$ is implemented in (a), whereas constant-time sampling of $C\alpha_i$ using φ2, g4*EA and $t_1$ are implemented in (b). The constant-time delay, T, was 14.3ms in (b), with ζ = 8 ms, Δ = 2*δ = 4.8 ms. Narrow and thick bars represent 90° and 180° high power pulses, respectively. Selective pulses on amide $^1$H were as follows: h1(90°) = 0.6 ms Sinc pulse at the selected $\nu$Hi; h2(180°) = 0.6 ms I3Snob pulse applied at 8.5 ppm (the center of the amide $^1$H region); and h3(180°) = 0.6 ms Sinc pulse at the selected $\nu$Hi. Selective $^{13}$C pulses were as follows. For the hncaCO experiment (a): c1(180° inversion) = 600 μs IBurp1 pulse at 54 ppm (the center of the Cα region); c2(90°) = 1000 μs Q5 pulse and c3(90°) = 1000 μs time reversed Q5 pulse at 173 ppm (the center of the CO region); c4(180° refocusing) = 700 μs Q3 pulse at 173 ppm; c5(180° inversion) = 600 μs I3Snob pulse at 173 ppm. For the hncoCA experiment (b): c1(180° inversion) = 600 μs I3Snob pulse at 173 ppm (the center of CO region); c2(90°) = 500 μs Q5 pulse and c3(90°) = 500 μs time reversed Q5 pulse at 54 ppm (the center of the Cα region); c4(180° refocusing) = 180 μs Q3 pulse at 54 ppm; c5(180° inversion) = 600 μs IBurp1 pulse at 54 ppm. Parameters for the continuous-wave blocks Hcw1, Hcw2, Ncw1, Ncw2, Ccw1, Ccw2, Ccw4, Ccw5 are summarized in **Table S1**. For (a), all gradient pulses have smoothed square (SMSQ) shape and a duration of 1ms. The relative gradient strengths are g1 = 57%, g2 = 37%, g3 =49%, g4 = 20%, g5 = 44.1%, g6 = 67%, g7 = 47%, g8 = -60%, and g9 = 24%. For (b,) all gradient pulses have smoothed square (SMSQ) shape and a duration of 1ms, except g4, g5 and g9, which have durations of 0.5 ms. The relative gradient strengths are g1 = 57%, g2 = 37%, g3 =49%, g4 = 80%, g5 = 44.1%, g6 = 67%, g7 = 47%, and g9 = 24%, whereas 100% corresponds to 53.5 G/cm. The pulse phases (default = x) are the same for (a) and (b): φ1=y,-y; φ2=2(x), 2(-x); φ3=4(x), 4(-x); φ4=8(x), 8(-x); and φrec=4(x, -x, -x, x).